\documentclass[fleqn,usenatbib]{mnras}

\usepackage{newtxtext,newtxmath}
\usepackage[T1]{fontenc}
\usepackage{ae,aecompl}
\usepackage{graphicx}
\usepackage{amsmath}	
\usepackage{amssymb}	
\usepackage[english]{babel}
\usepackage[table,x11names]{xcolor}
\usepackage{subfig}
\usepackage{rotating}
\usepackage{pdflscape}
\usepackage{color,soul}
\usepackage{afterpage}
\usepackage{hyperref}

\newcommand{\kms}{km~s$^{-1}$}
\newcommand{\msun}{{\it M}$_{\odot}$}

\newcommand{\be}{\begin{equation}}
\newcommand{\ee}{\end{equation}}

\newcommand{\hi}{H\,{\sc i}}
\newcommand{\mhi}{$M_{\mathrm{H\textsc{i}}}$}
\newcommand{\mstar}{{\it M}$_*$}

\title[HIRB-I]{The HI Morphology and Stellar Properties of Strongly Barred Galaxies: support for bar quenching in massive spirals}

\author[Newnham et al.]{
L. Newnham,$^{1}$\thanks{E-mail: lucy.newnham@port.ac.uk}
Kelley M. Hess,$^{2,3}$
Karen L. Masters,$^{4,1}$ 
Sandor Kruk,$^{5\thanks{ESA Research Fellow}}$
\newauthor Samantha J. Penny$,^{1}$
Tim Lingard,$^{1}$
R. J. Smethurst, $^{6}$
\\
$^{1}$Institute of Cosmology \& Gravitation, University of Portsmouth, Dennis Sciama Building, Portsmouth PO1 3FX, UK\\
$^{2}$Kapteyn Astronomical Institute, University of Groningen, Landleven 12, 9747 AD, Groningen, The Netherlands\\
$^{3}$ASTRON, the Netherlands Institute for Radio Astronomy, Postbus 2, 7990 AA, Dwingeloo, The Netherlands\\
$^{4}$Department of Physics and Astronomy, Haverford College, 370 Lancaster Ave, Haverford, PA 19041, USA\\
$^{5}$European Space Agency, ESTEC, Keplerlaan 1, PO Box 299, 2200AG Noordwijk, The Netherlands\\
$^{6}$Oxford Astrophysics, Denys Wilkinson Building, Keble Road, Oxford OX1 3RH, UK\\
}

\date{Accepted XXX. Received YYY; in original form ZZZ}
\pubyear{2018}

\begin{document}
\label{firstpage}
\pagerange{\pageref{firstpage}--\pageref{lastpage}}
\maketitle

\begin{abstract}
Galactic bars are able to affect the evolution of galaxies by redistributing gas in galaxy, possibly contributing to the cessation of star formation. Several recent works point to `bar quenching' playing an important role in massive disk galaxies like our own Milky Way. 

We construct the largest ever sample of gas rich and strongly barred disc galaxies with resolved HI observations making use of both the Giant Meter Radio Telescope (GMRT) and the Karl Jansky Very Large Array (VLA) to collect data. This sample of galaxies, which we call HIRB (HI Rich Barred) galaxies, were identified with the help of Galaxy Zoo - to find galaxies hosting a strong bar, and the Arecibo Legacy Fast Arecibo L-band Feed Array (ALFALFA) blind HI survey- to identify galaxies with an high HI content. 

We measure the gas fractions, HI morphology and kinematics in each galaxy, and use archival optical data from the Sloan Digital Sky Survey (SDSS) to reveal star-formation histories and bar properties. The HIRB galaxies presented here support a picture in which bar quenching is playing, or will soon play an important role in their evolution. They also support models which show how the presence of cold gas delays and slows the development of a strong bar. The galaxies with the lowest gas fractions (still very high for their mass) show clear HI holes, dynamical advanced bars and low star formation rates, while those with the highest gas fractions show little impact from their bar on the HI morphology, and are still actively star-forming. How such unusual galaxies came to be is an open question. Several of the HIRBs have local gas rich companions. Tidal interactions with these lower mass galaxies could result in an early triggering of the bar and/or accretion of HI between them. The role of environment in the evolution of the HIRB galaxies will be explored in a future paper. 
\end{abstract}


\section{Introduction}

Within a galaxy, a bar is a structure in the central region made up of stars, gas and dust following elongated orbits. Bars are found in 30-70\% of disc galaxies within 0.01 < z < 0.1 \citep[e.g.][]{MenendezDelmestre2007,barazza2008,nair2010a,Masters2011a}. A stellar bar has been demonstrated to disrupt and influence the motion of the material in the central region of galaxy \citep{Pfenniger1990,Regan1999}, as well as having an impact on the spiral arms \citep{Schwarz1984,Hart2018}; these impacts have been seen in many simulations \citep{OstrikerPeebles1973,Hohl1976,Athanassoula2002,Martel2013,Fanali2015, Carles2016, Spinoso2017, Khoperskov2018}, and are thought to affect the disc galaxy's overall evolution \citep{KormendyKennicutt2004,Athanassoula2012}, and since this process is slow, it's one of the processes of so-called ``secular" (or slow) evolution of galaxies.

The formation of a bar is not always a certainty in disc galaxies; after all not all spiral galaxies host bars (\citealt{Masters2011a,nair2010a}; a fact which even today is discussed as a puzzle as numerical simulations almost ubiquitously form them \citealt{SahaElmegreen2018}). The time it takes to develop a bar is thought to vary depending on a variety of dynamical factors in a galaxy, such as the ratio of stellar mass to halo mass, the gas content, rotational velocity etc \citep[e.g. for an extensive discussion see][]{Athanassoula2012}. Once the bar has formed, it will continue to grow in length and width with age, however simulations demonstrate that the correlation between the age of a bar and their length is not linear in a given galaxy, nor is the relation the same between different galaxies \citep[e.g.][]{Athanassoula2012}. Thus we cannot assume the age of a bar by looking at its physical length, or the length with respect to the diameter of the stellar disc, even though longer bars are likely to be dynamically older.  

Many studies have investigated which galaxies are more likely to have bars. In the largest recent studies, it was observed that disc galaxies that have a larger mass and are redder in colour are the most common hosts of bars \citep[e.g.][]{nair2010a,Masters2011a,CervantesSodi2017}. This correlation suggests that bars form around when a galaxy ceases to be star forming and is making the transition from the blue cloud to the red sequence. However, in \citet{NairAbraham2010}, the distribution is actually double peaked, with bars in both low mass and highest mass galaxies. It has also been observed that late type disc galaxies which have no bulge or at least a low bulge-to-total luminosity ratio are significantly more likely to host a bar \citep{barazza2008} and other results suggest no strong correlations \citep{Erwin2018}. These results may appear contradictory, but depend on details such as how the disc galaxies were selected (i.e. if redder spirals are excluded by a colour selection), and what the definition of a bar is (strong bars only, or also including weak bars). 

It has been observed that an increasing bar fraction in a galaxy population coincides with the increase in the average time since the start of star-formation quenching \citep{Masters2011a,Skibba2012,Smethurst2015}, an observation which suggests bars have some responsibility for quenching within a galaxy.  Additionally, there is evidence showing, that through funneling gas toward the centre, bars may be the cause of morphological quenching \citep{Athanassoula1992,Sheth2005,Masters2012,Cheung2013,Vera2016}. There is even a suggestion this process may have been important in the Milky Way's history \citep{Haywood2016}. 

A galaxy's environment has also been linked to its likelihood of developing a bar, with bars appearing more common in higher density regions. These, presumably tidally induced bars, may form and grow earlier than expected (related to the stellar mass of the host galaxy; \citealt{Noguchi1988,Giuricin1993,Moore1996,Eskridge2000,Skibba2012}). This is still debated, as some research finds no such link \citep{vandenBergh2002,Li2009,Marinova2012}, and some studies find only a poor link \citep{Lin2014}.

There exists a clear trend between galaxies that host strong bars and the atomic gas content \citep{DavoustContini2004,Masters2012,CervantesSodi2017}. Using the global neutral hydrogen (HI) measurements made by the Arecibo Legacy Fast Arecibo L-band Feed Array (ALFALFA) Survey \citep{Giovanelli2005,Haynes2011,Haynes2018}, \citet{Masters2012} found that galaxies hosting strong bars (identified with Galaxy Zoo as in  \citealt{Masters2011a}), are more likely to be gas-poor. \citet{Masters2012} also observed that galaxies that did not follow that trend, i.e. the ones that had a strong bar but were also gas-rich, were typically redder than similar gas-rich but non-barred galaxies, suggesting that star formation might have ceased earlier in those galaxies. 

Galaxies use their HI reserves to form stars \citep{Leroy2008,Saintonge2011}.  In dense regions, and usually in the presence of dust, HI will form H$_2$ in molecular clouds, which then cool further, collapse, and forms new stars.  Eventually the HI will run out in the galaxy and star formation will cease. Star formation may also cease if the HI is unable to cool due to significant torque or other dynamical processes \citep[e.g. bar torques][]{Tubbs1982}. 
The general picture that HI rich galaxies are star forming holds well, with a clear correlation between star formation properties and HI content \citep[e.g][]{Saintonge2012}. 

Numerical simulations of bar formation in disc galaxies can now realistically include a dissipative gas counterpart, and the effects of that on the galaxy. The formation of a bar in these is shown to aid the movement of material; with angular momentum being transported outwards and material transported in, towards the centre, in the inner bar regions \citep{BournaudCombes2002}. The gas is especially influenced by this since it is dynamically cold, it responds to gravitational perturbations \citep[e.g. as explored in][]{Athanassoula2012}

\citet{Kraljic2012}, exploring zoomed-in cosmological simulations and \citet{Athanassoula2013} uses detailed N-body and Smooth Particle Hydrodynamics (SPH) simulation to show the formation and growth of a bar with hydrodynamic models of gas flow, and present gas morphology as well as stars. In particular, \citet{Athanassoula2013} and \citet{Spinoso2017} predict the HI kinematics of strongly barred galaxies, the gas within the bar cororation radius (bar region) infalls to the very {\bf centre} (inner kpc) where it enhances star formation. Gas outside of this region is also prevented from entering due to the bar, therefore creating a `hole'. The gas is prevented from entering the central region as it cannot cross past co-rotation when the strong bar has been established. The gas is also transfered along the bar towards the centre. This agrees extremely well with the THINGS observations of M95 \citep{Walter2008} which show a large HI hole in the bar region. In this paper we will further test the predictions of these simulations in a sample of strongly barred gas-rich galaxies.

We introduce a sample of galaxies we are calling ``HI Rich Barred'' (HIRB) galaxies. This study was initiated to help reveal the evolution of galaxies with strong bars and large quantities of neutral hydrogen (HI). As mentioned above, galaxies tend not to host a strong bar and have a large reserve of HI \citep{Masters2012}. This is likely because the presence of significant quantities of gas inhibits and slows bar formation \citep{Athanassoula2013,Villa-VargasShlosmanHeller2010}. The HIRB galaxies are unusual as they all host a strong bar and still have large reserves of HI, such that they would be considered HI rich galaxies relative to typical scaling relations \citep{Huang2012}. 

The paper is organised as follows: in Section \ref{sec:Data} we discuss the data collected for this study, a description of the sample selection criteria, the available archival optical and spectroscopic data, and finally the resolved HI data collected for them by us using the VLA (NSF's Karl G. Jansky Very Large Array\footnote{The National Radio Astronomy Observatory is a facility of the National Science Foundation operated under cooperative agreement by Associated Universities, Inc.}) and the GMRT (the Giant Metrewave Radio Telescope in Pune, India). In Section \ref{sec:Results} we discuss the properties of the six HIRB galaxies revealed by these data. In Section \ref{sec:Discussion} we compare our results to simulations, and provide a plausible physical picture of the galaxies. We conclude and summarize our results in Section \ref{sec:Conclusions}. Where distances are needed to calculate physical lengths or other properties we use $H_0=$70 km/s/Mpc to convert from redshift. 

\section{Data}
\label{sec:Data}
\subsection{Sample selection}
\label{subsec:Sample selection and galaxy properties}

\begin{figure}
	\centering
	\includegraphics[width=0.50\textwidth]{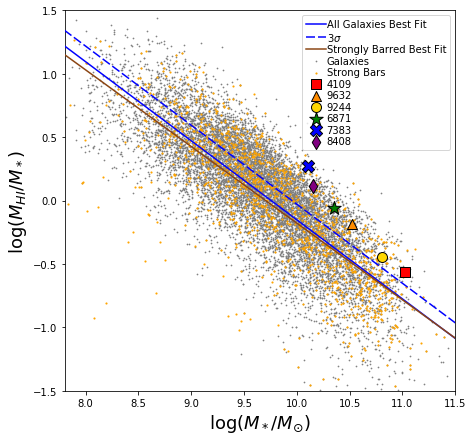} 
	\caption{HI content (from ALFALFA; \citealt{Haynes2018}), expressed relative to the stellar mass versus the stellar mass of galaxies in our sample (See Section \ref{subsec:Sample selection and galaxy properties}). A volume limited subset of ALFLAFA100 galaxies are plotted in grey and the orange points show all galaxies within that sample that host a strong bar ($p_{\rm bar}>0.5$). The blue line fits the trend between the HI mass fraction and the stellar mass, with the black dotted line showing a trend 3$\sigma$ more gas rich than that. The individually coloured shapes represent the six HIRB galaxies observed and detailed in this paper - the rainbow colour order reveals the absolute bar length (see Section \ref{subsec:OpticalMorphology}).  \vspace{-0.1in} }
	\label{fig:HImassfrac}
\end{figure} 

We introduce here the ``HI-Rich Barred'' (HIRB) galaxy sample. This sample has been designed to investigate a rare type of disc galaxy which are both strongly barred and very HI gas-rich. The parent sample consisted of 2090 almost face-on disc galaxies (selected through the Sloan Digital Sky Survey (SDSS)), with HI gas content detections from the ALFALFA 40\% survey \citep[which was the part of ALFALFA available at the time the HIRB galaxy survey was initiated][]{Haynes2011} and bar identifications from Galaxy Zoo \citep[GZ2][]{Willett2013} that were the basis of the study published in \citep{Masters2012}. This sample was built from a volume limited subset of the Galaxy Zoo sample, with $0.01<z<0.05$ (in order to allow for reliable distance measurements and an adequate angular resolution to detect bars in the galaxies). The SDSS spectroscopic limiting magnitude of $r$ = 17.7 corresponds to a volume limited sample of galaxies with $M_r$ < -19.0 out to $z$ = 0.05.

For more details of the selection of the parent sample from Galaxy Zoo and ALFALFA see \citep{Masters2012}. We show in Figure \ref{fig:HImassfrac} HI gas fraction from ALFALFA \citep{Haynes2011} against an estimate of the stellar mass, from K-correction fit to Petrosian magnitudes, for the entire sample (grey points) with strongly barred galaxies hi-lighted in orange. Solid lines show the best fit relation for all (blue) and just strongly barred (orange) galaxies, while the dashed line is 3$\sigma$ above this mean line. Our HIRB galaxies are shown by the large coloured stars, the colour of which, in rainbow order reveals a ranking by the physical size of the bar (which we discuss later in Section 3.1.1). 

Using this large sample we started by selecting objects which fit the criteria to be a HIRB galaxy, namely HI-rich and strongly barred. In \citet{Masters2012} it was shown that these are rare. We define the gas-rich sample to have a gas fraction, $\log$(\mhi$/$\mstar) ~greater than 3 $\sigma$ more than the mean value for their stellar mass as shown in Figure \ref{fig:HImassfrac}. We find that only 2\% (48 galaxies) of the entire parent sample meet both these criteria.  These 48 galaxies equate to 17\% of those galaxies with comparable HI mass for their stellar mass. 

Since these strongly barred, gas-rich galaxies are rare, it is hard to find a large sample of them nearby. We have to go to larger volumes (and therefore higher redshifts) in order to have a reasonable sample.  However, to be able to observe the HI at high enough resolution to resolve the bar region and with the column density sensitivity required to detect the disk requires tens of hours of observing for a single source at z>0.02. Due to this we ended up observing just seven of the nearest and/or largest in angular size sources for a total of 118 hours of telescope time on the GMRT and VLA combined. Unfortunately the data from one source was corrupted by a nearby continuum source (see \ref{subsec:ResolvedHIData} for details on the observing and data reduction). 

\begin{figure*}
	\centering
	\includegraphics[width=0.85\textwidth]{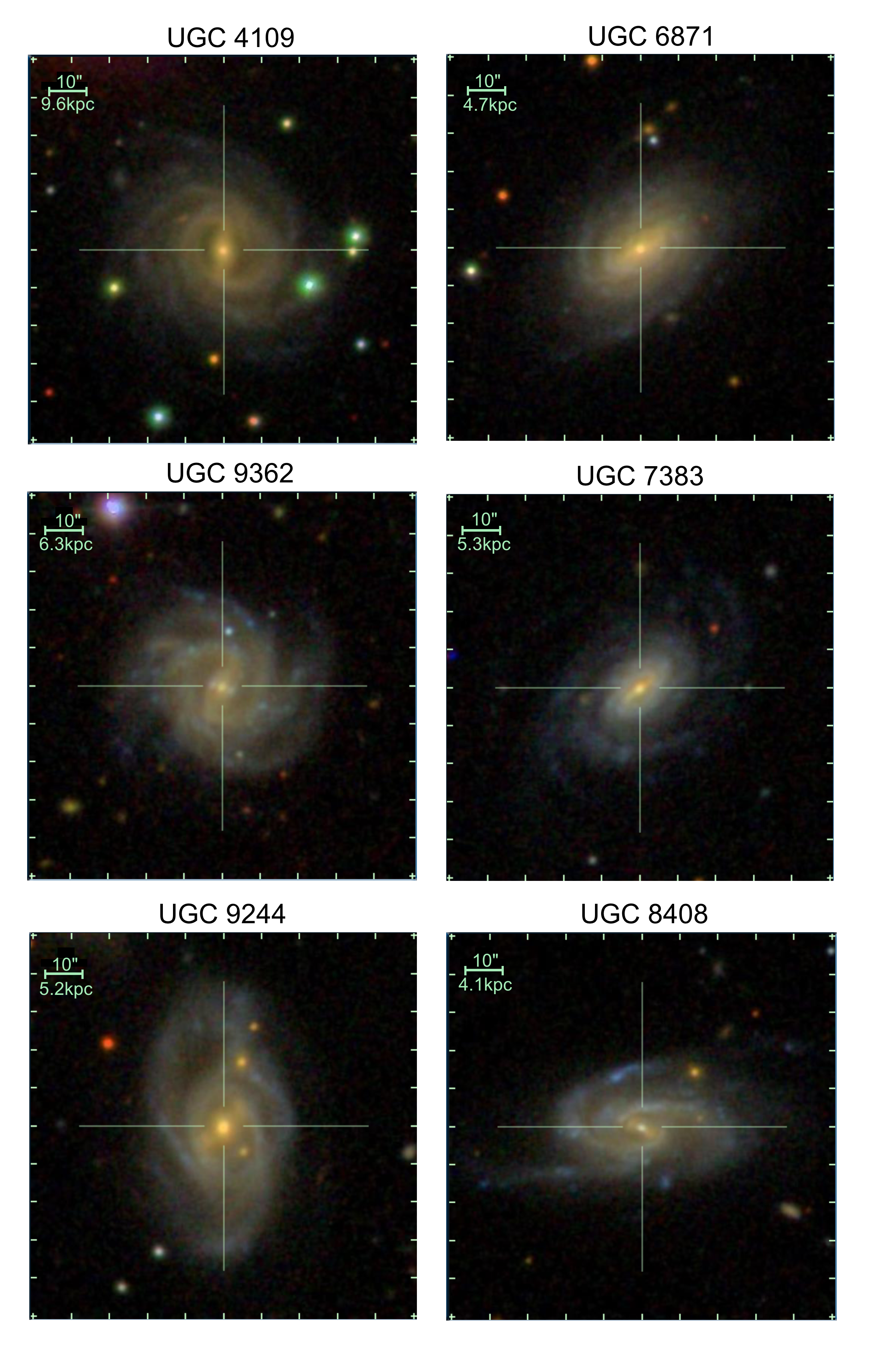} 
	\caption{Optical SDSS gri images of each of the six galaxies in the HIRB sample. \vspace{-0.1in} }
	\label{fig:all_sdss_1}
\end{figure*}

The properties of our final six galaxies are summarized in Table \ref{table:HIRB-selection}, where you can see they still span a range of various properties, while all being gas rich and strongly barred. SDSS optical $gri$ images are shown for all six in Figure \ref{fig:all_sdss_1}. 
 
\begin{table*}
	\centering
	\caption{Properties of the seven HIRB galaxies used in sample selection. Column 4 is the debiased fraction of GZ users identifying a bar, 7 is the global width from ALFALFA \citep{Haynes2011} (8) Shape and colour used to represent in plots
    \label{table:HIRB-selection}}
	\begin{tabular}{lcccccccc}
		\hline
        \hline
		Name & R.A. and Dec & $z$ & $p_{\rm bar}$ &  $\log (M_\star$& $\log(M_{\rm HI} $ & W50 & Shape and Colour\\
	  	& (J2000) &                  &                          &   $/M_\odot)$   & $/M_\odot)$     &  \kms &	\\
		(1)   & (2)                 &   (3)       &  (4)             & (5)        & (6)                  & (7)    &(8)     \\
        \hline
		UGC 5830 & 10 42 38.0 +23 57 07&  0.044 & 0.62 & 11.17 & 10.40 & 624 &  \\
        UGC 4109 & 07 56 16.6 +11 39 41 & 0.046 & 0.91 & 11.04 & 10.46 & 304 & Red Square\\
        UGC 9362& 14 33 17.5 +03 54 10  & 0.030  & 0.75 & 10.50 &   10.34 & 113 & Orange Triangle\\
		UGC 9244 & 14 26 08.4 +05 14 15&  0.028  & 0.59  & 10.94 & 10.36 & 414 & Yellow Circle\\
		UGC 6871 & 11 53 46.3 +10 24 10 &  0.022 & 0.54 & 10.55 &10.29  & 346 & Green Star\\
		UGC 7383 & 12 20 01.3 +08 36 26  & 0.025 & 0.58 & 10.10 & 10.38 & 294 & Blue x\\
		UGC 8408 & 13 23 00.3 +13 57 02   & 0.024  & 0.88 & 10.19 & 10.26 & 290 & Purple Diamond\\

		\hline
	\end{tabular}
\end{table*}

Our HIRB galaxies share many properties with those in the (HI) HIghMass Survey ongoing at the JVLA \citep{Hallenbeck2014} in that they are unusually massive and gas-rich, and so like HIghMass, make good $z\sim0$ analogues of the types of galaxies which will dominate detections in SKA surveys (as well as the similar HI mass and gas-rich galaxies in HIGHz at $z\sim0.2$; \citealt{HIGHz}). HIghMass galaxies were selected without any reference to visual morphology, and there are no obvious strongly barred galaxies in that sample. The general conclusion from the five HIghMass galaxies with resolved HI is that they must be galaxies about to transition from a gas rich but relatively inactive state, to a state of vigorous star-formation \citep{Hallenbeck2014}.

There are other ongoing surveys of HI-rich galaxies which are also relevant for comparison. The Bluedisks project \citep{Wang2013} is using Westerbork Synthesis Radio Telescope (WSRT) observations to investigate the reason for excess HI in some galaxies over that predicted from scaling relations, while \citet{Lemonias2014} present VLA C configuration observations of a sample of HI rich, but not star-forming spirals (including for UGC 4109 in our sample). Meanwhile \citet{Lee2014}'s ``HI Monsters'' survey is a CO survey of very HI rich galaxies. None of these surveys use information on galaxy morphology for selection, nor comment on it in their results, while it is central to the HIRBs selection. In both Blue disks and HI Monsters, they find that HI rich (but not necessarily high HI fraction) spirals are simply scaled up versions of lower mass spirals, and sit on all the typical scaling relations. The passive HI rich spirals of \citet{Lemonias2014} however show more extended and lower surface brightness HI than is typical, and it is interesting to note that roughly half of them appear to host relatively strong bars (see their Figure 3; although we have not explored the significance of this fraction).

\subsection{Optical Photometry, Spectra and Morphologies}

All optical photometric and spectroscopic data used in this paper is taken from the final release of the first phase of the Sloan Digital Sky Survey (SDSS Data Release 7, \citealt{DR7}), and all HIRB galaxies (and their parent sample) are part of the Main Galaxy Sample of SDSS \citep[MGS,][]{Strauss2002}

We make use of morphologies from the second phase of Galaxy Zoo \citep[GZ2,][]{Willett2013}. These  come from the aggregation of classifications made by numerous volunteers on images from the SDSS for the brightest 25\% of MGS galaxies (those with $m_r<17$). In GZ2 a median of 45 people looking at each galaxy. The GZ2 classification process starts with the question: `Is this galaxy simply smooth and rounded, with no signs of a disc?', where one of the options were `features or disc', followed by `Could this be a disc viewed edge-on?'. The weighted and debiased fraction of users answering a specific way to a single question in GZ2 is denoted, $p_X$. The sample defined in Masters et al 2012 used $p_{\rm features}p_{\rm not~edge-on}$>0.25. After these questions, the citizen scientists were asked `Is there a sign of a bar feature throughout the centre of the galaxy?' From this we can get the probability of a galaxy having a bar. We call this $p_{\rm bar}$ and define this as the percentage of `yes' votes to that question (although note that it is not exactly this, following a correction for redshift bias as described in \citealt{Willett2013}). For example, a galaxy with $p_{\rm bar}$=0.5 suggests that 50\% of the people asked if there is a visible bar responded `yes'. A comparison with other bar idenfitications is shown in Appendix A of \citet[and also see \citealt{Willett2013}]{Masters2012}. Using this, $p_{\rm bar}$>0.5 is classified as a strong bar (SB) which is used for the HIRB galaxies. As discussed above, the parent sample is a subset of the Galaxy Zoo 2 sample which overlaps with the ALFALFA 40\% footprint. 

\begin{figure}
	\centering
	\includegraphics[width=0.50\textwidth]{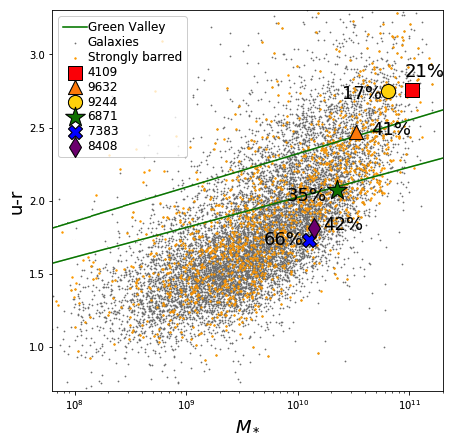} 
	\caption{A colour-mass diagram showing the volume limited ALFALFA100 galaxies (grey), those of which are strongly barred (orange) and with the HIRB galaxies hi-lighted with individual shapes and colours. Two of the HIRBs are at the high mass end of the blue cloud,  two are at opposing sides of the green valley (within the green lines \citep{Schawinski2014}) and the remaining two are in the red sequence. The percentage numbers adjacent to each HIRB galaxy displays the gas fraction that galaxy has.\vspace{-0.1in} }
	\label{fig:ColourMass}
\end{figure}

SDSS provided optical photometry in $ugriz$-bands, and with a variety of apertures. In the NASA Sloan Atlas \citet{Blanton2005} provide matched aperture photometry also including  NUV and FUV photometry from the GALEX satellite \citep{Martin2005}. We use the Absolute magnitude in rest-frame GALEX/SDSS FN$ugriz$, from Petrosian apertures to generate colours, and the stellar mass from K-correction fit to Petrosian magnitudes. A colour-mass diagram is shown for the parent sample in Figure \ref{fig:ColourMass}, with the six HIRB galaxies hi-lighted with the large coloured symbols. As is obvious from that diagram HIRB galaxies span a range of optical colours. 

The SDSS images can also be used to provide more quantitative details of the bar structures. Following the method described in \citet{Kruk2017} we obtained two or three component decompositions for each galaxy in the sample. This is particularly useful in providing bar lengths and bar position angles. We give more details on how bar lengths are obtained in Section \ref{subsec:bar_lengths} below. 

We further make use of the SDSS fiber spectra to obtain measurements related to the stellar populations and ionized gas in these galaxies. We note that the SDSS fibre has a 3'' size, which represents only 1.2--2.9 kpc at the distance of the HIRB galaxies, as such these spectra measure the properties of only the central parts of these galaxies, well within the bar radius. A full spectral fitting is beyond the scope of this work, however we make use of line fluxes and spectral index measurements provided by the Max Planck Institute for Astrophysics and John's Hopkins University collaboration (MPA-JHU)\footnote{http://www.mpa-garching.mpg.de/SDSS/}. These star formation rates are found using the technique described in \citet{Brinchmann2004} while updating the procedure for aperture corrections. This procedure included calculating the light outside the fiber per galaxy, then fitting stochastic models to the photometry, akin to those used in \citet{Salim2007}\footnote{https://wwwmpa.mpa-garching.mpg.de/SDSS/DR7/sfrs.html}.

\subsection{Resolved HI Data}
\label{subsec:ResolvedHIData}

Our HI observations were designed to resolve the optical bar at the center of the galaxy in addition to the rest of the disk.  However the scarcity of gas-rich, strongly barred galaxies meant that the majority of our targets are at relatively large distances ($0.022<z<0.046$) compared to typical HI interferometric targets ($z<0.02$).  The small beam required to resolve the centers of these galaxies, necessitated observing with the B-configuration of the VLA with its maximum baseline of 10 km.  The GMRT is comprised of 30 antennas similarly arranged in a fixed ``Y''-configuration with a maximum baseline of 25 km, but also has 14 antennas concentrated within 1 km.  The combination of long baselines with a compact core mean that the performance of the GMRT is often compared to the VLA in B-configuration.  In balancing resolution with column density sensitivity, our galaxies were challenging targets to observe.  In this section we describe our observing, imaging, and source finding strategies.  Table \ref{table:observations} summarizes the properties of the final data cubes for all six galaxies.

\subsubsection{VLA Data Reduction}
 
We observed UGC 5830, UGC 4109, UGC 6871, and UGC 7383 with the VLA in B-configuration between November 2013 - January 2014 (Project ID: 13B-142).  Each observation was centered on the \hi\ systemic velocity of the galaxy measured by ALFALFA.  The bandwidth spanned 8 MHz divided into 512 channels, resulting in a 3.3~\kms\ resolution over approximately 1700~\kms.  We observed each galaxy for 6x2 hours, including calibration, resulting in a total of 10 hours on-source.  The data were calibrated and imaged using standard methods in Common Astronomy Software Applications (CASA; \citealt{McMullin2007}).  The final images were made with natural weighting to provide the best column density sensitivity.

We imaged each 2 hour observing session individually to inspect the quality of the data before combining all sessions together.  However, UGC~7383 required special attention. Although we were able to image all the individual sessions and achieve the expected rms, $\sigma$, for unknown reasons the CASA \texttt{clean} task was unable to properly combine the first session with the remaining five. Instead, we imaged the first session individually, the last five sessions together, and combined the two resulting cubes in the image plane, weighting them by $1/\sigma^2$ to create a final dirty (not yet cleaned) image cube of the full 10 hours.  We performed the same weighting combination procedure on the synthesized dirty beam cubes.  In order to clean the dirty image cubes, we imported the data to the Astronomical Image Processing System (AIPS; \citealt{Greisen2003}) and cleaned the channels individually using the AIPS task \texttt{APCLN}.

For UGC~5830, we realized after imaging that a $>0.3$ Jy source lay at the half power point of the primary beam, causing significant difficulties for imaging and continuum subtraction.  We attempted to do self-calibration on the field, and to subtract the continuum with a number of higher order polynomials, but could not completely remove the interfering continuum source.  We concluded that ``third generation'' calibration techniques--which include direction dependent gain calibration--would be required, but given the faintness of \hi\ profile from ALFALFA (UGC 5830 has the second highest redshift in our sample), the effort to make a reliable \hi\ map outweighed the benefits.  In the rest of the paper, we exclude UCG~5830 from the analysis.

\subsubsection{GMRT Data Reduction}

We observed UGC~9244, UGC~9362, UGC~8408 with the GMRT for 22-24 hours each including calibration, with a total of approximately 18-19 hours on source. The observations took place in February-March 2016 and February 2017, while the galaxies were up at night to avoid solar interference and to minimize the incidence of other ground based radio frequency interference.  We observed using the GMRT's spectral zoom mode with 256 channels across 4 MHz bandwidth centered on the ALFALFA HI redshift of the galaxy.  This provided 3.3~km~s$^{-1}$ spectral resolution over approximately 800~km~s$^{-1}$.  Unfortunately, due to roll-off at the edges of the sideband, less of the band was useable.  In the worst case, combined with RFI, the final cube covered 460~km~s$^{-1}$, which fortunately included the galaxy and a few channels for continuum subtraction.

We calibrated the data in the standard way using AIPS while on site at the GMRT.  For imaging purposes the data were spectrally averaged as little as possible to be able to detect reliable signal in individual channels, while still resolving the motion of the gas in the disk.  The galaxies were imaged with natural weighting to provide the best column density sensitivity as possible given the long baselines.  In addition, UGC 9362 benefited from self calibration in AIPS which reduced the noise in the final image by approximately 1/3.

Despite careful flagging we found that there were still stripes in some of the images suggesting bad baselines which we had been unable to isolate from looking at the individual visibilities.  Stripes in the image plane are equivalent to a high point in the UV plane, and its complex conjugate which has the same value.  From the orientation and width of the stripes we can predict the location of the high points in the UV plane.  Thus, in order to eliminate the stripes, for each galaxy we Fourier transformed the image cube using the \texttt{fft} task in Miriad \citep{Sault2011}, we blanked the outlier points in the resulting UV plane data cubes (each image cube produces an amplitude and phase cube) and we performed the inverse Fourier transform to return to the image plane. (The inverse also produces an amplitude and phase cube but the phases of sky image are zero everywhere.)  In the end we eliminated the worst offending stripes, and reduced the noise in the cubes by a factor of approximately 1.055.

Finally, we spatially smoothed the data of UGC 6871 so the final spatial resolution and column density sensitivity was of comparable to those of the other galaxies (see Table \ref{table:observations}).

\subsubsection{Source Finding}

In order to create moment maps from the relatively low signal-to-noise data we used the automated \hi\ source finder SoFiA, the Source Finding Application \citep{Serra2015}.  We used the ``smooth and clip'' method with a large merge radius and pixel dilation to capture the large-scale low surface brightness flux in each galaxy.  The source finder works iteratively, to find sources in the data down to a user specified signal-to-noise, mask them, smooth the data and search again.  The mask is iterated upon and grows with each smoothing run.  We used kernels which smoothed the data up to only 2 channels in velocity, since our cubes were already averaged significantly in velocity, and up to $12\times12$ or $24\times24$ pixels spatially.  This allowed us to identify and include low column density gas in our final images, while the moment maps are presented at full spatial resolution to resolve the central regions of each galaxy.

\begin{table*}
	\centering
	\caption{
    Properties of the HIRB galaxy final data cubes. Columns are (2) The spatial resolution of the final image, (4) The rms of the final image, (5) The resolution of the restoring clean beam.}
    \label{table:observations}
	\begin{tabular}{ccccccc}
		\hline
        \hline
		Galaxy & Spectral Resolution & Robustness  &  rms             & Resolution & HI Column Density    & Observatory \\
        Name   & (\kms)                & (Weighting) &  (Jy~beam$^{-1}$~chan$^{-1}$) &  (arcsec)  & atoms cm$^{-2}$~\kms &             \\
         (1)   & (2)                 &   (3)       &  (4)             & (5)        & (6)                  & (7)  \\
		\hline
		UGC 4109 & 19.8 & natural & $1.8\times10^{-04}$ & $8.9\times5.1$  & $1.1\times10^{20}$ & VLA  \\ 
		UGC 9362 & 31.9 & natural & $8.0\times10^{-05}$ & $7.3\times4.8$  & $1.0\times10^{20}$ & GMRT \\ 
        UGC 9244 & 27.5 & natural & $2.3\times10^{-04}$ & $7.8\times5.3$  & $2.3\times10^{20}$ & GMRT \\ 
		UGC 6871 & 19.8 & natural & $3.1\times10^{-04}$ & $5.9\times4.8$  & $3.0\times10^{20}$ & VLA  \\
		"        & "    & smoothed & $5.5\times10^{-04}$ & $10.1\times8.2$  & $1.8\times10^{20}$ & "  \\ 
		UGC 7383 &  9.9 & natural & $3.9\times10^{-04}$ & $10.9\times5.1$ & $8.5\times10^{19}$ & VLA  \\
        UGC 8408 & 27.5 & natural & $2.1\times10^{-04}$ & $9.6\times4.8$  & $1.8\times10^{20}$ & GMRT \\
	\hline
	\end{tabular}
\end{table*}

\section{Results}
\label{sec:Results}

\subsection{Optical Properties of HIRB galaxies}
\subsubsection{Bar Lengths}
\label{subsec:bar_lengths}
The length of a bar is often difficult to define and there is no standard way of measuring it. In this study, we use bar length or refer to twice the bar radius. Commonly used techniques to measure bar sizes are based on (1) visual bar length measurements (for example drawing a line on top of the bar, used in Galaxy Zoo studies, \citep{Hoyle2011}, (2)  bar major axis surface brightness profiles (obtained from photometric decompositions, \citet{Kruk2018}) and (3) ellipse fits to galaxy isophotes (where the bar length is assumed to be at the maximum of the ellipticity, \citet{Sheth2003,Erwin2005}).

We have used all the three methods to determine the bar sizes for the six galaxies in the HIRB Galaxy Survey. First, we measured the bar lengths using visual estimation directly from the SDSS $i$-band images (which are less prone to dust, and hence better probe the mass distribution). In the second method, the bar size is assumed to be the effective radius of a S\'ersic fit to the brightness profile along the bar major axis, in a detailed disc+bar+bulge photometric decomposition done in \citet{Kruk2018} The multi-component decomposition also allows the position angle of the bar and of the disc to be measured. Finally, we used the \texttt{iraf} ellipse routine to fit elliptical isophotes to the galaxy $i$-band images and determined the bar size at the maximum of the ellipticity. The bar length measurements agree (within $\sim$10\% of each other). In general, the bar effective radius is shorter than the manually measure bar length, which is shorter than the estimated value based on maximum ellipticity. Henceforth, we use the visually measured bar lengths, which are the often the median value of the three measurement methods, while the position angles are based on the disc+bar+bulge decompositions.

While bar length cannot be used as any kind of absolute chronometer for bar age, all simulations show monotonic increase in bar length with time (e.g. \citealt{Athanassoula2012,Martinez-Valpuesta2006}). We therefore use absolute bar length (measured in kpc) as a proxy for bar age, and tag the HIRB galaxies in order of their bar length as revealed by the colour coding of their symbols in all plots (see Figures \ref{fig:HImassfrac}, \ref{fig:ColourMass}, \ref{fig:sf-seq}, \ref{fig:dn4000} and \ref{fig:starpy}). 

\begin{table}
	\centering
	\caption{Optical Properties of the HIRB galaxy sample Columns are (3) Diameter of the disc (4) Bar to Galaxy ratio, (5) Optical $(u-r)$}
    \label{table:optical}
	\begin{tabular}{lcccc}
		\hline
        \hline
		Name & Bar length &  $2r_{\rm Petro}$ & Relative bar &  $(u-r)$  \\
             &  (kpc) &  (kpc) &  length  &     \\
             (1)   & (2)            &   (3)       &  (4)      & (5)          \\
		\hline
		UGC 4109 & 14.48 & 48.26 & 0.29 & 2.76 \\
		UGC 9362 & 11.5 & 30.33 & 0.36 & 2.47 \\
        UGC 9244 & 8.22 & 28.60 & 0.28 & 2.75 \\
		UGC 6871 & 6.36 & 17.55 & 0.36 & 2.08 \\
        UGC 7383 & 5.12 & 14.97 & 0.30 & 1.73 \\
		UGC 8408 & 4.26 & 20.85 & 0.21 & 1.81 \\
		\hline
	\end{tabular}
\end{table}%

\subsubsection{Optical Morphology}
\label{subsec:OpticalMorphology}

In this section we provide some discussion and comments on the optical morphology of each of the six HIRB galaxies. 

\begin{itemize}
\item {\bf UGC 4109:}
 is a flocculent spiral which was classified as SB(r)b in \citep[][hereafter RC3]{Corwin1994}, denoting an Sb type with a strong bar and ring. In GZ2, 41/45 classifiers identified the bar, and four out of five users asked also noted the ring (questions about rings in Galaxy Zoo are in a submenu, often missed). According to our measurements, this galaxy has the longest bar of all of the galaxies in the HIRB sample (with a length of 14.5 kpc), which stretches across 29\% of the optical extent of the galaxy. This galaxy is also one of the most massive of the sample, with a stellar mass of $10^{11.0}$\msun,  and one of the `reddest' of the HIRBs. This galaxy is shown as the red square icon in all sample plots. 

\item {\bf UGC 9362:}
like UGC4109, UGC 9362 is a flocculent spiral galaxy. It was classified as Sbc in the RC3 who somehow missed the very obvious bar on photographic plates; 21/28 GZ2 users identified the obvious bar. It has the second longest bar of the HIRB sample, with a length of 11.5 kpc which however covers 36\% of the total extent of it's optical disc (i.e. more of the galaxy than the bar in UGC 4109). This, which is the third most massive in the HIRB sample is optically found in the green valley (close to the red side) and is shown as the orange triangle icon in all sample plots. 

\item {\bf UGC 9244:}
this galaxy has some obvious interaction with a very close neighbour to the North-East. This has resulted in the northern most spiral arm being significantly distorted. UGC 9244 has two primary arms and a third smaller one. It was classifed as SBbc in the RC3. It hosts the third longest bar of the sample at 8.2 kpc, or 28\% of the diameter of the disc. It has a stellar mass of $10^{10.9}$\msun, making it the second most massive galaxy of the sample. It is also another of the `redder' galaxies in the sample. This galaxy is shown as the yellow circle icon in all sample plots. 

\item {\bf UGC 6871:}
is central in the range of optical colours and stellar masses of the HIRB sample. It has a mass of $10^{10.6}$ \msun~ and appears to be on the blue side of the green valley. Although UGC 6871 is clearly another flocculent spiral, there is a distortion in one of the arms at the lower left of the galaxy. It appears to be stretching out from the galaxy towards the end of the arm. Curiously the RC3 classifies this as an SB0 (perhaps due to the limitations of photographic images blurring the flocculent arms together), while \citet{nair2010a} classify it as Sc, missing the fairly obvious bar. In GZ2, 21 out of 39 classifiers indicated the presence of the bar. The bar is the third shortest of the sample at 6.4 kpc or 36\% of the diameter of the disc. This galaxy is shown as the green star icon in all sample plots. 

\item {\bf UGC 7383:}
is an obviously blue galaxy. It also has a stellar mass of $10^{10.1}$\msun, making it the least massive of the sample. Accordingly, UGC 7383 has one of the physically shortest bars of the sample at 5.1 kpc, which however still covers 30\% of the extent of the optical disc. The optical image of this galaxy shows two defined, tightly wound spiral arms, without much evidence of interaction with any nearby galaxies. The RC3 classifies this as Sab, again missing the obvious bar, spotted by 22 out of 38 GZ2 classifiers. This galaxy is shown as the blue x icon in all sample plots. 

\item {\bf UGC 8408:} 
is also known as NGC 5115. This is the final HIRB galaxy and has the shortest actual bar length at 4.26 kpc, which is less than third of the length of the longest bar in the sample. While UGC 8408's bar still covers 21\% of the disc's optical extent, this is the shortest relative bar length of the entire sample, which hints that it might be the most recently formed. 

UGC 8408 is one of the `bluest' galaxies we observed, it is also the second least massive galaxy with a stellar mass value of $10^{10.2}$\msun. There is some obvious distortion to the optical morphology of this galaxy. It appears to be a grand design spiral, with two main arms which are clearly defined. However it seems to also have an extra `arm' which does not quite reach the centre. 

In the RC3 this galaxy was classified as SBcd. This galaxy is shown as the purple diamond icon in all sample plots. 
\end{itemize}

\subsubsection{Star Formation History}
\label{subsec:SFproperties}

\begin{figure}
	\centering
	\includegraphics[width=0.50\textwidth]{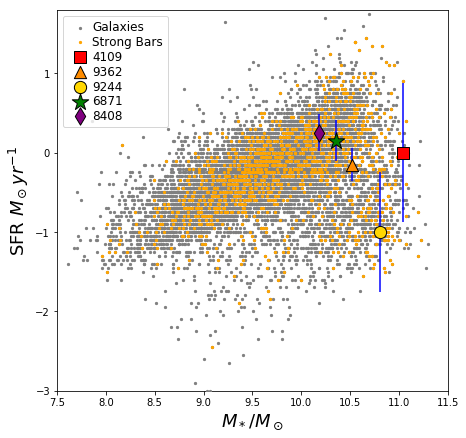} 
	\caption{The HIRB galaxies plotted on a Star Formation Rate vs stellar mass (a galaxy star forming sequence) plot. The grey points show our volume limited subset of ALFALFA100, while the orange points depict the strongly barred of those galaxies, and the coloured symbols show the HIRB galaxies, with the rainbow order indicating physical bar length.  \vspace{-0.1in} }
	\label{fig:sf-seq}
\end{figure}

The HIRB sample span a range of properties including gas fraction (Figure \ref{fig:HImassfrac}), galaxy total color (Figure \ref{fig:ColourMass}), and star formation rate (Figure \ref{fig:sf-seq}), despite covering a relatively narrow range in stellar mass (approximately one order of magnitude). From Fig \ref{fig:ColourMass} we see that our galaxies span the blue cloud, green valley and red sequence. In the following sections we present the optical properties and HI morphologies in relation to the bar length to attempt to explain how they vary with the transition of galaxies with a high HI content from the blue cloud to the red sequence.

We explore the star formation properties of the sample further in Figure \ref{fig:sf-seq} which makes use of star formation rates from the MPA-JHU catalog. These star formation rates are used to plot the ``star forming sequence'' of galaxies. We show the entire parent sample (ALFALFA detected GZ2 galaxies) in grey, with those with strong bars overplotted in orange. Five of the HIRB galaxies are hi-lighted, the final HIRB galaxy (UGC 7383) has an unreliable flag on its SFR, and so is not plotted, but its optical colours and spectrum suggest it is star forming.  We see that only two are clearly on the star forming sequence, with two just below, and one well off it. This is despite all five having significantly more HI than is expected for their stellar mass. 

While the SF rates shown in Figure \ref{fig:sf-seq} have been corrected for H$\alpha$ emission from non- star forming sources it is reasonable to wonder if the presence of the strong bar might be feeding an AGN in these galaxies through the in-falling gas. This was shown as likely by \citet{OhOhYi2012}, but disputed by \citet{Galloway2015} and \citet{Cheung2015}. According to a diagnostic ``BPT'' diagram \citep{Baldwin1981}, all six of the HIRB galaxies and their companions have line emission which comes almost exclusively from star formation regions. None of these galaxies present any evidence for any emission from an AGN. 

\begin{figure*}
	\centering
	\includegraphics[width=1\textwidth]{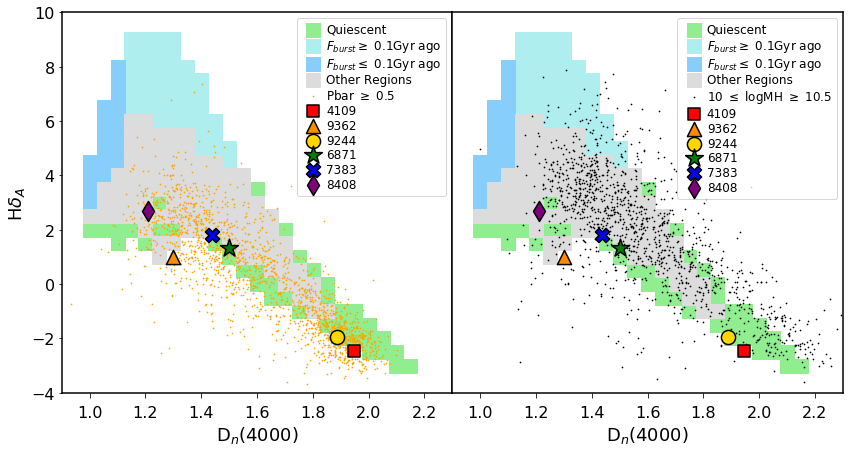}
	\caption{D$_n(4000)$ versus H$\delta_A$ for: Left: all spiral galaxies with a strong bar ($P_{bar}$). Right: spiral galaxies with a HI mass range similar to that shown by the HIRB's (log(\msun/\mhi) = 10-10.5). D$_n(4000)$ correlates with the global stellar population age (larger is older), while H$\delta_A$ peaks in post-starburst galaxies. The background colours shows models of \citet{Kauffmann2003a}; green for quiescent galaxies, cyan for starburst galaxies (burst occurred less that 1 Gyr ago), blue shows post-starburst galaxies (burst occurred over 1 Gyr ago) and grey are all other regions covered by the model galaxies \citet{Kauffmann2003a}. These values are from the 3'' SDSS fiber, so we probe the stellar population in the bulge. As with the other figures, the individual HIRB galaxies are shown by different shapes and colours. \vspace{-0.1in} }
	\label{fig:dn4000}
\end{figure*}

Another useful diagnostic diagram for the optical properties of galaxies $D_n(4000)$ against $H\delta_A$. This diagram is diagnostic of the fraction of star formation in the last 2 Gyrs which occured in bursts, versus continuous star formation \citep{Kauffmann2003a}. We show this in Figure \ref{fig:dn4000}, which compares the position of the HIRB galaxies with similar HI masses (left) or similar bar properties (right). Models from \citet{Kauffmann2003a} are indicated, which show regions which are populated by smooth star formation histories (green), currently starbursting (cyan) or post-starburst (blue) galaxies, with emission from galaxies with mixed star formation found in the middle. As is noted, the HIRBs typically lie in the quiescent part of this diagram, as is also normal for galaxies with similar bar properties (at right), but lie to the low star formation side of galaxies with similar HI masses. 

As will be discussed in Section \ref{subsec:HIMorphology}, UGC 4109 (red square) and UGC 9244 (yellow circle) both have HI holes at their centre. This suggests the bar has already funneled the gas towards the very centre, which might then lead to a concentration of H$_2$ in the centre and therefore we expect younger population of stars in the centre (e.g. as was observed by \citealt{Ellison2011} in a sample of barred galaxies).  

The HIRB galaxies were run through the star formation history software tool \textsc{starpy} \citep{Smethurst2015}. \textsc{starpy} employs a Bayesian method along with ensemble Markov Chain Monte Carlo \citep{Foreman-Mackey2013} to infer the parameters describing a simple exponentially declining star formation history of a single galaxy. \textsc{starpy} makes use of the SDSS and GALEX optical photometry, specifically the Petrosian magnitude \texttt{petroMag} $u$ and $r$ wavebands, provided by SDSS Data Release 7 \citep{Stoughton2002} and the $NUV$ waveband from GALEX \citep{Martin2005}. \textsc{starpy} requires the observed $u-r$ and $NUV-u$ colours and redshift. Intrinsic dust is not taken into consideration or modeled for. The full description and method can be found in Section 3.2 in \citet{Smethurst2015}. 

\begin{figure}
	\centering
	\includegraphics[width=0.5\textwidth]{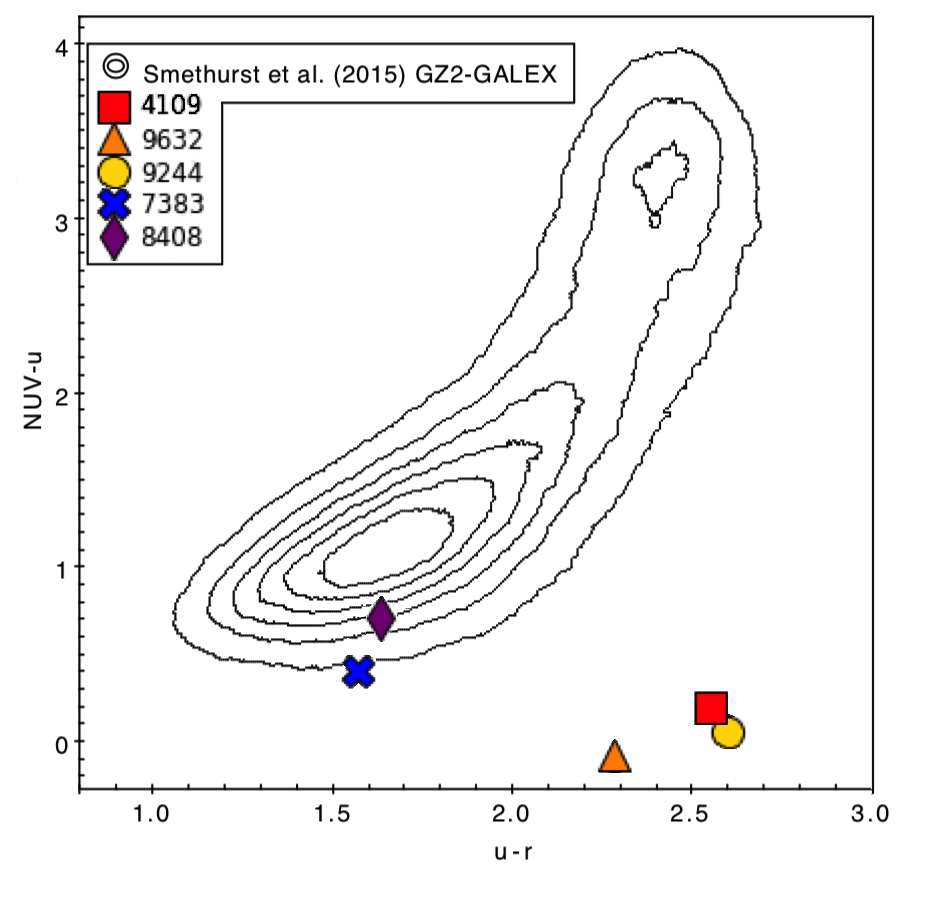}
	\caption{A $NUV-U$ vs $u-r$ plot for the 5 galaxies that were run through \textsc{starpy}. Compared to the \citet{Smethurst2015} sample, 3 HIRB galaxies (UGC 4109, UGC 9362 and UGC 9244) are very optically red in comparison to their blue NUV colours. \vspace{-0.1in} }
	\label{fig:starpy}
\end{figure}

Only five of the HIRB galaxies were able to go through the \textsc{starpy} software. UGC 6871 was not detected with GALEX; it was right on the edge of a field, meaning that the $NUV$ value was not available. Three of the HIRBs (UGC 4109, UGC 9244 and UGC 9362) have results which suggested they started to quench very early on in their history, and the quenching proceeded slowly so they are probably still star forming. The other two (UGC 8408 and UGC 7383) have begun quenching more recently and relatively rapidly. Figure \ref{fig:starpy} shows the HIRB galaxies placement in comparison with the sample used in \citet{Smethurst2015}, where we can see that two HIRB galaxies are relatively normal $u-r$ colours, while three are very red in $u-r$ for their $NUV-u$ colours indicating the extremely slow quenching.

While the star-formation properties and stellar ages revealed by the optical data on HIRB galaxies do not show a monotonic increase with bar length, there is clearly a general trend such that those HIRB galaxies with longer bars (red, orange and yellow) are more likely to be passive and have old stellar populations for their stellar mass than those with shorter bars (blue and purple). This is particularly evident in Figure \ref{fig:starpy} where the two groups show quite different quenching histories. 

\subsection{HI Properties of HIRB Galaxies}
\begin{table*}
	\centering
	\caption{Our measurements of HI properties of HIRB Galaxies using the GMRT/VLA. Columns are (2) central velocity of HI detection, (3) HI mass }
    \label{table:radio}
	\begin{tabular}{lcccc}
		\hline
        \hline
		Name & Central $v_{HI}$ & $\log{M_{\mathrm{H\textsc{i}}}/M_\odot}$ & \# of HI &  HI Hole?\\
		&      &      & companions                    &          \\
                 (1)   & (2)                 &   (3)       &  (4)             & (5)        \\
		\hline
		UGC 4109 & 13087 & 10.62 & 0 & Offset Hole \\
		UGC 9362 & 8528 & 10.00 & 1 & Hole \\
		UGC 9244 & 8106 & 10.11 & 1 & Hole \\
		UGC 6871 & 6460 & 10.03 & 0 & Offset Hole \\
		UGC 7383 & 7177 & 10.59 & 2 & No Hole \\
		UGC 8408 & 7093 & 10.07 & 0 & No Hole \\
		\hline
	\end{tabular}
\end{table*}%

\subsubsection{HI Masses from Synthesis Imaging}
\label{subsec:HIMasses}

We measure the HI mass for each galaxy in the HIRB sample using the interferometers (VLA and GMRT), and ALFALFA had already measured the HI mass using the single dish telescope, Arecibo. The masses we measured differ by small amounts to the masses from ALFALFA, but this is to be expected. ALFALFA's \mhi ~is shown in Table \ref{table:HIRB-selection}, column 6, and the \mhi measured from the resolved HI data can be seen in Table \ref{table:radio}, column 3. ALFALFA's measurement was larger in four of the cases (UGC 9362, UGC 9244, UGC 6871, UGC 8408). In the other two, where ALFALFA's HI measurement was lower than that collected by the interferometer (UGC 4109 \& UGC 7383), the data was collected both times by the VLA rather than the GMRT.

In all cases we believe the ALFALFA mass is more likely to be a good measurement of the total HI mass. Single dish radio observations will collect and measure all of the HI detected within the beam, which is significantly larger than the beam of interferometers like the VLA and the GMRT. This means the measured \mhi value may  include some of flux from HI in other nearby galaxies or structures at are too nearby to differentiate between. While reducing resolved HI data, we can disregard some of the HI we can see is coming from other structures, however synthesis imaging will also resolve out some of the low surface brightness signal received. Because of this, synthesis imaging is known to miss HI mass, and it is common practise for the single dish data to be treated as more reliable measure of total HI mass. We have determined that the any nearby galaxies to the HIRB galaxies are outside the Arecibo beam. 

\subsubsection{HI Morphology and Velocity Field}
\label{subsec:HIMorphology}

In this section we will describe the HI morphology and content from our VLA/GMRT observations. According to simulations \citep{Athanassoula2012}, there should exist a correlation between gas fraction, the rate of development of a strong bar, the bar's formation time, and the morphology of the HI remaining in the galaxy. In this section we investigate if this correlation is also present from the data collected by the HIRBs survey. 

\begin{figure*}
	\centering
	\includegraphics[height=0.75\textheight,angle=270]{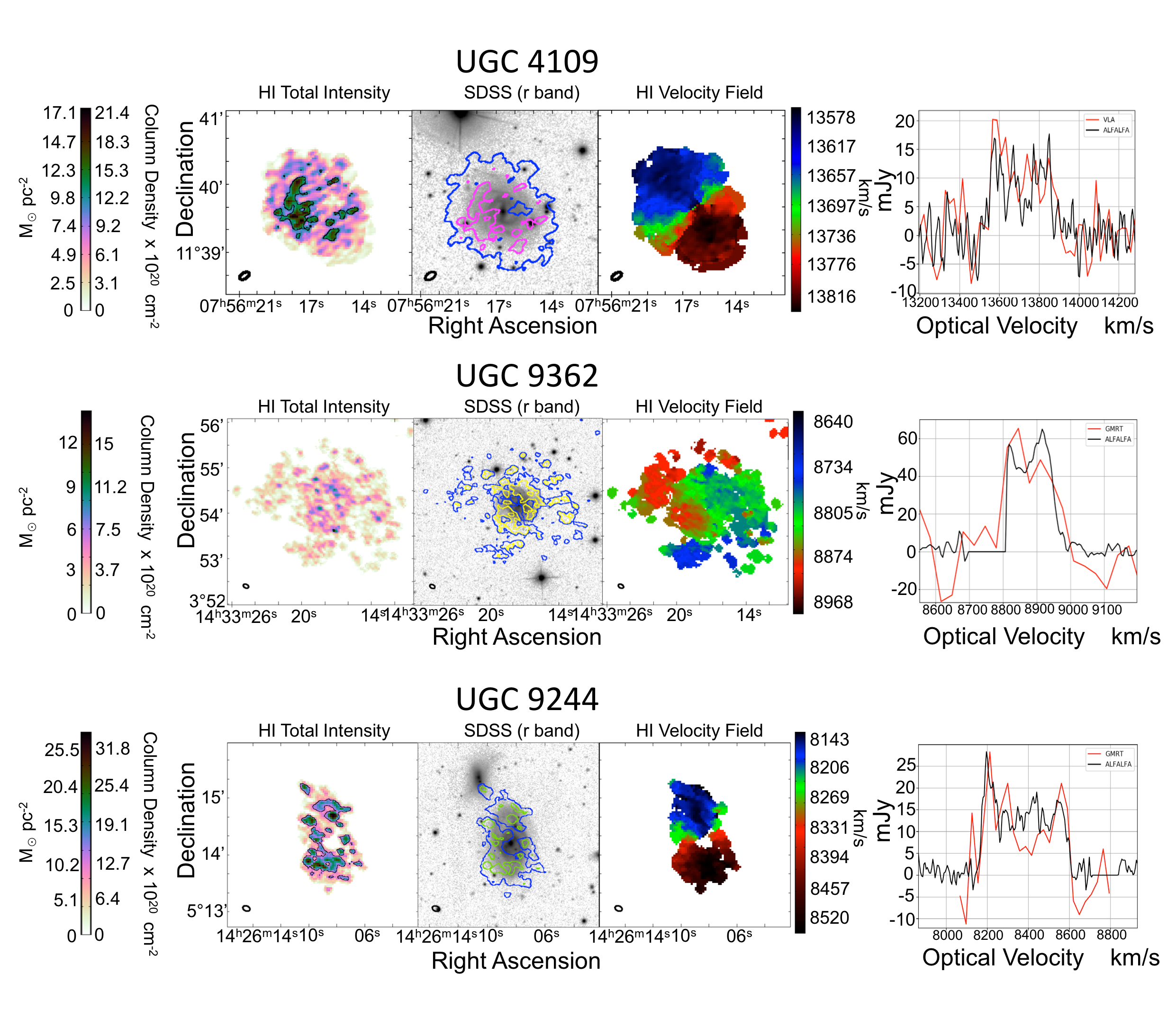} 
	\caption{\scriptsize HI resolved Data visualized for UGC 4109, UGC 9362 and UGC 9244. For each galaxy the panels from left to right are; (1) The HI total intensity map depiciting the column density of each pixel. The black contour depicts the star formation threshold at 10 \msun pc$^{-2}$ \citep{Schaye2004}, (2) The SDSS r band optical image of the galaxy in greyscale with the contours of the HI total intensity map overlayed. Contour colours are; blue = 3$\sigma$, yellow = 5$\sigma$, green = 8$\sigma$, magenta = 10$\sigma$, cyan = 20$\sigma$, orange = 30$\sigma$, red = 50$\sigma$. (3) The HI velocity field of the galaxy, expressed in km/s. (4) A comparison of the HI `double peak' detected originally by ALFALFA (black) and by our observation with the VLA/GMRT (red).  \vspace{-0.1in} }
	\label{fig:casa1-3}
\end{figure*}

\begin{figure*}
	\centering
	\includegraphics[height=0.75\textheight,angle=270]{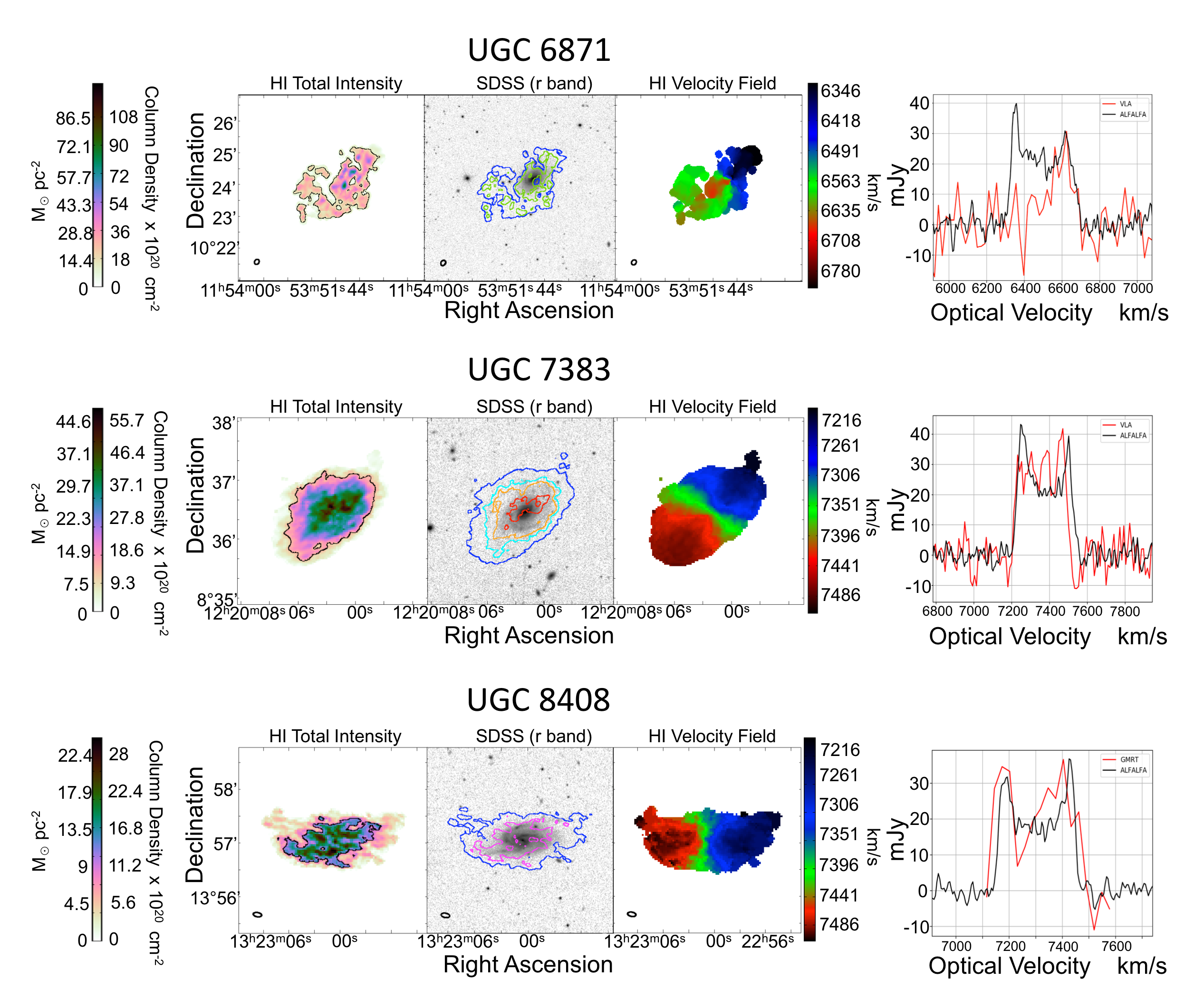} 
	\caption{\scriptsize HI resolved Data visualized for UGC 6871, UGC 7383 and UGC 8408. For each galaxy the panels from left to right are; (1) The HI total intensity map depiciting the column density of each pixel. The black contour depicts the star formation threshold at 10 \msun pc$^{-2}$ \citep{Schaye2004}, (2) The SDSS r band optical image of the galaxy in greyscale with the contours of the HI total intensity map overlayed. Contour colours are; blue = 3$\sigma$, yellow = 5$\sigma$, green = 8$\sigma$, magenta = 10$\sigma$, cyan = 20$\sigma$, orange = 30$\sigma$, red = 50$\sigma$. (3) The HI velocity field of the galaxy, expressed in km/s. (4) A comparison of the HI `double peak' detected originally by ALFALFA (black) and by our observation with the VLA/GMRT (red).  \vspace{-0.1in} }
	\label{fig:casa4-6}
\end{figure*}

The HI gas fractions (\mhi/(\mhi+\mstar)) for our six galaxies are shown in Figure \ref{fig:ColourMass}, and the morphologies shown for each HIRB galaxy in Figures \ref{fig:casa1-3} and \ref{fig:casa4-6}. We also summarize this data in Table \ref{table:radio}. 

\begin{figure*}
	\centering
	\includegraphics[width=0.7\textwidth]{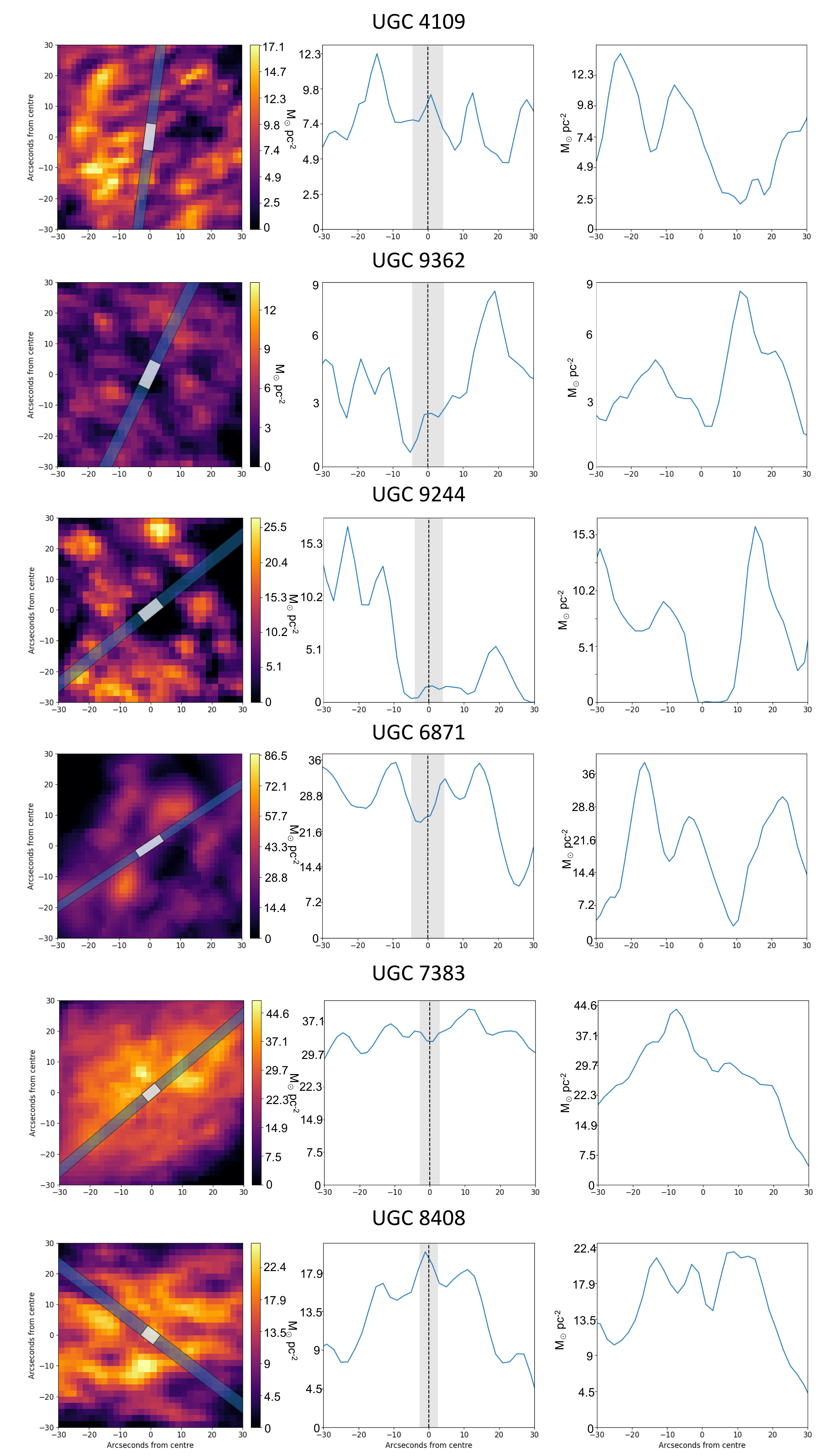} 
	\caption{Left: The HI density map of each galaxy. A slice from the data is then taken along the bar and the HI density is then plotted (middle). The area of the bar is highlighted. Right: A slice from the data is taken perpendicular to the bar and the HI density is plotted.\vspace{-0.1in} }
	\label{fig:all_slice}
\end{figure*} 

It is clear from Figures \ref{fig:casa1-3} and \ref{fig:casa4-6} that there is a variety of HI morphology revealed by the HIRB galaxies. Figure \ref{fig:all_slice} shows the intensity of HI in a slice aligned at the position angle of the bar; the highlighted region shows exactly where the bar is to aid in our description of how HI morphology relates to the position of the bar. 

Both UGC 9362 and UGC 9244 show clear large scale holes in their centres and Figure \ref{fig:all_slice} shows the absence of any siginificant amounts of HI in their bar regions at all.  

While UGC 4109, and UGC 6871 show large scale HI holes, they appear to be significantly offset from their centres. However when looking at the highlighted region in Figure \ref{fig:all_slice} we can see that there is still a dip in both of these galaxies' HI intensity at exact position of the bar. This could suggest that the bar is working towards sweeping the central region, but this still does not offer an explanation for the source of the large offset hole. It may be that tidal disruption plays a role in the HI morphology shown in these two galaxies.  

UGC 4109 has also been observed by the VLA in C-configuration in \citet{Lemonias2014}. Named GASS 51390 in that study, they show the HI intensity map in their Figure 6. While this C configuation has lower spatial resolution than our data, we can clearly see there exists a hole in HI. Our observations reveal that the hole is slightly offset from the centre. On the opposite side to the offest hole there is a denser region - it is possible that a minor merger has occured resulting in the unusual HI morphology. 

UGC 7383 and UGC 8408 are the HIRB galaxies with the two lowest bar lengths, and neither appear to have a hole in the centre of their HI. Looking at the intensity along the exact section of the bar, there is still no evidence of any significant dip in the HI intensity. UGC 8408 on the other hand does have a dip, albeit small enough to not notice when looking solely at the intensity map. So perhaps the bar is working to clear the middle of gas after all.

Our VLA observations of UGC 6871 did not detect all the gas associated with the blue-shifted side of the galaxy. This is evident from the HI spectrum in Figure \ref{fig:casa4-6} (top row, fourth column), in which the emission which is blue shifted with respect to the systemic velocity of the galaxy is seen in the ALFALFA spectrum, but missing from the VLA spectrum. Examining the HI intensity map overlaid on the optical image (Figure \ref{fig:casa4-6}), we see that the blue-shifted side of the HI disk (northwest) is not as extended compared to the stellar disk, as the red-shifted side of the gas disk is. Initially we thought this asymmetry was tidal in origin, however given the missing flux in the HI profile, we suspect that the gas may be present, but at a column density below what we detect in the VLA observations.

In the first panel for each galaxy we overlay a conservative threshold for the HI density typically associated with star formation (10 \msun pc$^{-2}$ shown by the black contour). In UGC 4109  the majority of this higher density HI is found to the left of the hole in the HI, in a dense region. UGC 9244 and UGC 6871 reach the star formation threshold in many places in the HI however their bar region HI holes are quite evident. Our two blue galaxies with the shortest bars, UGC 7383 and UGC 8408 have HI above the star formation threshold throughout their discs, which fits with the picture of them forming stars throughout. Finally UGC 9362 is the only one of the six HIRB galaxies which does not reach this star forming in any significant area (there is a small part which does around RA: 14h33m18.5s, Dec: 3\textdegree 54' but it is too small to be of any significance morphologically). This galaxy does however appear to be forming stars, which must be happening at a lower HI surface density than is typical. 

HI observations also provide velocity fields. we can see that the velocity field for four of the HIRB galaxies are rotating regularly (UGC 4109, UGC 9244, UGC 7383 and UGC 8408). The remaining two galaxy's velocities (UGC 9362 and UGC 6871) appear to be somewhat distorted. In UGC 9362, the `bluer' velocities are distributed through the middle velocities (green) rather than collected on one side. Similarly, UGC 6871's `redder' velocities are concentrated in the middle of the galaxy rather than being on one side showing rotation. There is a clear disruption in the gas's regular rotation. It is possibly that the source of this distortion is nearby companions to these galaxies; this will be discussed further in a future paper (L. Newnham et al. in prep.)

\section{Discussion}
\label{sec:Discussion}

The HIRB study was initiated with the goal of investigating what a strong bar does to a gas rich galaxy, or equivalently what a significant quantity of gas in a galaxy does to the evolution of a strong bar. 

Bar quenching is an idea which has been around in the literature for some time (e.g. \citealt{Tubbs1982}) but has recently gained more attention due to the observation that passive spirals have large bar fractions \citep{Masters2010}, the strong trend of bar fraction with optical colour \citep{Masters2011a,nair2010a}, and suggestions that bars can drive secular growth of bulges, a morphological characteristic long agreed to correlate with quenched star formation (e.g. \citealt{Cheung2013,Kruk2018}). Simulations of the impact of a bar on gas in a galaxy, demonstrate its ability to redistribute gas (e.g. \citealt{Berentzen1998,Athanassoula2013,Combes2008,Villa-VargasShlosmanHeller2010}), more recently simulations have explicitly explored the possibility for these flows to lead to global SF quenching \citep{Gavazzi2015,JamesPercival2016,Spinoso2016,Khoperskov2018,JamesPercival2018}. However there is also a possibility that the torques from the bar simply prevent gas from collapsing and forming stars at the typical thresholds for star-formation (e.g. the model initially proposed by \citealt{Tubbs1982}). 

In \citet{Athanassoula2012} a galaxy's gas fraction was found to be influential in the rate of development of the bar itself, with higher gas fractions delaying the formation of the bar, and resulting in weaker bar. Evidence of this link between gas fraction and the suppression of bar formation is present in observations which show that bars are less likely to be found in gas rich spirals (or equivalently that barred galaxies have lower gas content;  \citealt{Masters2012,Kim2017}). 

We seek to use observations of HI morphology to provide constraints on the mechanism of bar quenching, and to understand how unusual galaxies with high gas fractions and strong bars came to be. If gas redistribution is the main mechanism we might expect to see clear evidence for it in the HI morphology (e.g. central holes), while if HI gas is present in large quantities and above the typical threshold for star formation ($\sim$3-10\msun pc$^{-2}$, \citealt{Schaye2004}) in galaxies where star formation is suppressed, we can argue that bar torques preventing the gas from cool may be more important. 

Gas-rich and strongly barred galaxies have not been studied extensively with resolved HI observations, partly because most surveys for resolved HI do not consider morphology in their selection, so are likely to miss strongly barred galaxies, and while there are a small number of very nearby galaxies with strong bars and resolved HI observations these tend not to be particularly gas rich galaxies. However there are resolved HI observations of very nearby galaxies with strong bars. 

The HI Nearby Galaxy Survey (THINGS) \citep{Walter2008} survey reaches typical column densities of 4 x 10$^{19}$ cm$^{-2}$, which is comparable to the column densities we reach with the HIRB sample. They present observations for one of HIRB-like strongly-barred galaxy, M95, showing in extreme detail the distribution of the HI throughout the galaxy revealing a HI hole (and a small HI concentration in the very centre was also found).  Recently, \citet{George2018} took these data, along with other multi-wavelength data for M95 to argue that redistribution of gas is likely to be the main process for bar quenching in M95 (rather than heating from bar torques).

The proto-typical strongly barred galaxy, NGC 1300, also has resolved HI data, taken in the 1980s with the VLA \citep{England1989}. A re-analysis of those data show clearly the HI hole and gas streaming \citep{Lindblad1997}. A large HI hole is also observed in nearby barred galaxies NGC 3992 \citep{Bottema2002}, and NGC 7479 \citep{Laine1998}, and while there is clear hints of a HI hole in the date on NGC4123 presented by \citet{Wiener2001} they also note evidence for shocks caused by gas flow in the bar region. Contrary to this, the HI morphology in the very late-type barred spiral, NGC 3319 does not show a central hole, but rather gas and star formation all along its bar \citep{Moore1998}. This difference is explained by \citet{Moore1998}, as NGC 3319 being a dynamically younger galaxy than the early-type spirals with strong bars, obvious HI holes and lower gas fractions. Indeed it has a significantly higher gas fraction that the other galaxies (56\% compared to 6\% gas in M95). It is worth noting that all of these very nearby galaxies with resolved HI have typical (or even low) HI gas fractions for their stellar mass, unlike our HIRB sample which are all at least 3$\sigma$ gas-rich outliers in the HI mass to stellar mass relation (see Figure \ref{fig:HImassfrac}, where are all HIRB galaxies are well above the mean line, while these galaxies (not shown) would be on or below it).

We will now discuss how the data presented in this paper for HIRB galaxies support, or reject the plausible bar quenching mechanisms of gas redistribution and/or heating due to bar torques in very gas rich galaxies. 

We particularly compare our observations to the simulations of \citet{Athanassoula2012}. In that work it is shown that the time taken to develop a HI hole will vary from galaxy to galaxy, and also depends on the details of the gas fraction in the galaxy, with observable HI holes being slow to develop in gas rich galaxies even in the presence of a strong bar. In that work, the most gas rich galaxies did not develop clear HI holes by the end of the simulation (at 10 Gyrs). It is also known that the bar mass impacts the ability of a bar to evacuate HI, with \citet{Hunter1990} showing that a bar needs to hold about 10\% of the disc mass to make a central HI depression. 

We start by now considering the sample in order of HI gas fraction and compare to the gas morphology and timescales shown in Figure 4 of \citet{Athanassoula2012}: 

\begin{itemize}
\item {\bf Gas Fraction $\sim$20\%}:
From the simulations of \citet{Athanassoula2012}, all strongly barred galaxies that have a total gas fraction around 20\% develop the hole in the centre of the HI by 6 Gyr after the formation of a strong bar. The HIRB sample's two most gas poor galaxies UGC 4109 and UGC 9244 with HI gas fractions of 21\% and 17\% show an offset hole, and central hole respectively (Figure \ref{fig:ColourMass}, and Figure \ref{fig:casa1-3}). These galaxies are therefore consistent with being at this stage of their evolution where they formed a bar at least 6 Gyr ago. These are also our two globally reddest galaxies, suggesting a global cessation of star formation, this is backed up, for the central bulge region at least, by their fibre spectra which reveal old central stellar populations (Figure \ref{fig:dn4000}) , and further while UGC 4109 (with an offset hole) is only just below the star forming sequence, UGC 9244 which has the largest central HI hole is well below the star forming sequence (see Figure \ref{fig:sf-seq}). M95, with a gas fraction of 6\% by mass \citep{Leroy2008} appears to be another example of this type of dynamical advanced galaxy with a strong bar. 

\item {\bf Gas Fraction $\sim$35-45\%}: \\
Three of our galaxies have HI gas fractions around 35-45\%. One of these galaxies is in the redder edge of the green valley, one in the bluer edge and one in the blue sequence, as seen in Figure \ref{fig:ColourMass}, with UGC 9362 (orange triangle) on the red side, UGC 6871 (green star) on the blue side and UGC 8408 (purple diamond) well in the blue cloud. 

According to the simulations of \citep{Athanassoula2013}, a galaxy with 50\% of its disc mass in cold gas and a bar which formed at least 6 Gyr ago would already have a noticeable hole in the centre. Indeed both UGC 9362 (orange triangle) and UGC 6871 (green star) have central HI holes (although UGC 6871's is offset from the centre), however UGC 8408 (purple diamond) does not. This suggests the bar in UGC 8408 may be dynamically younger than those in both UGC 9362 and UCG6871, and UGC8408's more vigorous star-formation and younger stellar population paints a similar picture of a dynamically younger galaxy. 

\item {\bf Gas Fraction $\sim$70\%}: \\
Our most gas rich galaxy, UGC 7383, despite having a stellar mass of $M_\star \sim 10^{10}$\msun, where gas fractions of around 30-40\% are more typical, has a gas fraction of 66\% (Figure \ref{fig:ColourMass}) and displays no evidence for a HI hole in the centre. However looking at the most gas rich galaxies simulated in \citet{Athanassoula2013} we would not expect to see a HI hole in this type of galaxy unless the bar formed more than 10 Gyr ago, as galaxies with these high gas fractions were not observed to form a HI hole before the end of the simulation.
\end{itemize}

Overall our observations support a picture where bar formation is delayed, and therefore the development of a corresponding HI hole is delayed in galaxies with very high gas fractions, but that bars are apparently acting to clear HI holes in the centres of galaxies and therefore accelerate the cessation of star-formation globally. 

We will now look at how the star formation properties and optical morphology of the galaxies correlate with the HI morphology. If the development of HI holes are driving bar quenching in HIRB galaxies we'd expect that those with holes are under star forming relative to typical galaxies with their properties, while those without holes may be more normal. 

 In fact most of the galaxies in the HIRB sample appear to be relatively quiescent, with only two of them appearing in the middle of the star forming sequence (Figure \ref{fig:sf-seq}), despite their large HI reserves. They also show a range of global optical colours. 

\begin{itemize}
\item {\bf Dynamically Old Bars (HI Hole)}

Two of the HIRBs show obvious central HI Holes (UGC 9362, orange triangle, and UGC 9244, yellow circle; Figure \ref{fig:all_slice}), while two have apparently offset HI holes (UGC 4109, red square and UGC 6871, green star). These four have the longest physical bars again supporting a picture that their bars are dynamically the oldest among the HIRB galaxies. 

They do however show a range of star formation properties, with UGC 9244 (yellow circle) well off the star formation sequence, UGC 9362 (orange triangle) and UGC 4109 (red square) at the lower edge and UGC 6871 (green star) appearing to still be forming stars at a typical rate. Looking at $Dn4000$ as a tracer of stellar population age we again see two galaxies with clearly old populations (UGC 4109, red square and UGC 9244, yellow circle), while the other two are younger. Finally, the analysis of likely quenching history using the technique of \citet{Smethurst2015} reveals that the three galaxies with HI holes where this could be done (UGC 6871 did not have a GALEX observation) show a clear signature of unusually slow quenching (red $u-r$ colours, while remaining relatively blue in $NUV-u$; see Figure \ref{fig:starpy}). 

All in all this supports the idea that a HI hole correlates with secular (slow) quenching of star formation, possibly caused by bar driving gas clearing, however the details of the individual galaxies are, perhaps not surprisingly hinting at a more complex picture. 

We will come back to the role of local environment in more detail in a future paper (L. Newnham et al. in prep.), but it's clear from the HI morphology, and detections of companions in the wider field of our data that local interactions may play a role in the history of these galaxies. For example, looking at the HI distribution in UGC 9244 (Figure \ref{fig:casa1-3}, bottom row, first panel) we can see that it is clearly interacting with a lower mass companion. The HI gas is extended in the direction of the companion, distorting the gas from an elliptical shape. This corresponds with the distortion of the spiral arm, visible in the optical image of this galaxy (Figure \ref{fig:all_sdss_1}, lower right. There is also evidence of interaction in the HI morphology of UGC 9362 with it's nearby companions, NSA 18045 (Figure \ref{fig:casa1-3}, middle row, first panel). Although the data has not picked up any evidence of a HI bridge between the two, we can see that the HI in NSA 18045 is heavily concentrated in the side of the galaxy closest to UGC 9362. 

We have not commented on a mechanism which might create a offset HI hole not centred on the bar, or galaxy (as see in both UGC 4109 and UGC 6871). This may again point to some tidal distortions caused by interactions, and perhaps gas rich low mass companions are the reason for HIRBs having such high gas fractions relative to their other properties. Curiously though, UGC 4109 has no companions which appear local enough to this galaxy to account for it's distortion, and its HI velocity field is very regular, so that explanation seems unlikely in this case. The other HIRB galaxy with an offset HI hole, UGC 6871 is also fairly isolated and while it's HI velocity field looks quite disturbed, we think the VLA data may have missed some lower column density HI on the blue-shifted side which was detected in the single dish data. 

\item {\bf Dynamically Young Bar (No HI Hole)}

Both of our HIRB galaxies with no evidence for HI holes (UGC 7383; blue x, and UGC 8408; purple diamond) also have the shorter physical bars (and the highest gas fractions) supporting a picture where the gas has delayed bar formation, and the bar has not yet had time (or perhaps is not massive enough) to sweep out cold gas in the bar region. 

These galaxies are both optically blue, and UGC 4808 is found on the normal star-forming sequence (UGC 7383 has an unreliable flag in the MPA-JHU derived star formation so we cannot comment on it's relative location on that diagram), they also both have relatively young population ages, and in the models of \citet{Smethurst2015} appear on the main locus of points with evidence perhaps for recent quenching. The bar which has not swept out a HI hole in either of these, perhaps does not create enough heating to prevent star formation in them either.  
\end{itemize}

Despite their higher than average HI content (at least 3$\sigma$ above the average HI content for their stellar mass as shown in Figure \ref{fig:HImassfrac}), not all of the HIRB galaxies are actively star forming. Our observations of the HI morphology of HIRB galaxies support a picture where as a bar forms, grows and develops it sweeps out a hole in the HI gas. The central gas will be displaced quicker the lower the gas fraction of the galaxy, i.e UGC 4109 and UGC 9244 both have a low gas fraction and have evidence of a hole at their centre (UGC9244) or offset slightly (UGC 4109). So for those two galaxies the bar didn't necessarily have to age significantly before the HI morphology had been disrupted. This also fits with our two galaxies with the shortest bars, neither UGC 7383 or UGC 8408 have obvious holes in the centre and happen to be the galaxies with the highest gas fractions in the sample. This suggests that bars need more time to grow to be able to sweep out the gas from the centre. We can see a slight indent in the intensity of the HI in the centre of these two galaxies (Figure \ref{fig:all_slice}) which might hint that the bar is staring to clear the gas. As seen by their range of optical colours (Figure \ref{fig:ColourMass}) and the inconsistencies with the star formation trends (Figure \ref{fig:sf-seq}), these galaxies are in the transition from being star forming and blue, becoming red and ceasing to form stars due to the presence of the bar. 

\section{Conclusions}
\label{sec:Conclusions}

We have collected, reduced and imaged resolved HI data for 6 galaxies making up the HI Rich Barred (HIRB) galaxy survey. The galaxies were selected using the ALFALFA survey and GZ2 morphologies, using the parent sample defined in \citet{Masters2012}, and found to have a HI mass in the range of 10.25 < \mhi/\msun < 10.47. These values correspond to extremely high gas fractions for galaxies of these stellar masses, with the gas fracton (\mhi/\mstar) more than 3$\sigma$ above the average for their \mstar. Strongly barred galaxies were selected using the GZ2 bar ``probability" of $p_{\rm bar} > 0.5$. Any value less than that would not be considered a strong bar, either a bar does not exist in that galaxy, or it is weak. This resulted in a sample of 48 galaxies which satisfied all of the criteria, we needed to reduce the number further, in order to observe and reduce the resolved HI data for them. We selected the nearest ten of these, six of which we were able to obtain HI total intensity and velocity fields for using with the GMRT or the VLA. These six galaxies have a range of optical colours and bulge sizes, making the sample representative of the local massive spiral population. 

The HIRB galaxies show a variety of star formation properties, despite the presence of a strong bar, which usually correlates with passive spirals \citep{Masters2010,Fraser-McKelvie2016} and the presence of large reserves of HI which usually correlates with strong star formation \citep{Saintonge2012}. While some galaxies have central fiber spectra which place them on on the star formation sequence of galaxies (Figure \ref{fig:sf-seq}), they are all in the quiescent region of the D$_n$4000 vs. H$\delta_A$ plot (Figure \ref{fig:dn4000}), although with a wide range of D$_n$4000 values, suggesting a wide range of population ages. 

Secondly we find that as predicted by the simulations of \citet{Athanassoula2013}, not all of the HIRB galaxies have a HI `hole' in the centre, where the gas has been driven into the centre or prevented from entering the region due to the strong bar. The simulations show that the timescale for HI hole formation depends on the gas content, so we explain this as some of the strong bars in the sample are in such gas rich galaxies that the bar formation has been delayed and they have not had adequate time to funnel all of the gas in the central region. 

We observe a correlation between the HI morphology and star formation properties of the galaxy, which are divided into those which have dynamically mature bars, HI holes and have apparently ceased star formation, and those which are more gas rich, do not have HI holes (suggesting dynamically younger bars) and are still actively star forming. This does support a picture in which bar quenching play an important role in the evolution of disc galaxies. 

 Through the resolved HI data we see that four of the HIRB galaxies have nearby companions; not all were obviously interacting from the SDSS optical images. This seemingly high proportion of interactions suggests a close `fly-by' or an impending merger of the galaxies, and this disruption to the secular evolution could have either tidally triggered the formation of the bar, or added HI gas to the galaxy after the bars had formed. This process might offer an explanation as to why those galaxies have formed their bar before they have had a chance to use all of the HI reserves. This role of environment on HIRB galaxy evolution will be investigated further in a future paper (Newnham et al. in prep.). 
 
 Galaxy evolution is a complex system in which many physical processes play a role. While the role that bars and other secular processes play in galaxy evolution is not always considered in studies of the general galaxy population, we demonstrate with our HIRB sample that it can be significant in massive spiral galaxies, and therefore it's role in the broader processes of galaxy evolution should not be neglected. Further study with larger samples of strongly barred galaxies with resolved gas and stellar kinematics will be needed to fully understand the significance of this process to the broader galaxy population. 

\paragraph*{ACKNOWLEDGEMENTS.} 

We thank the staff of the GMRT who have made these observations possible. GMRT is run by the National Centre for Radio Astrophysics of the Tata Institute of Fundamental Research. The Karl G. Jansky Very Large Array is operated by The National Radio Astronomy Observatory. AIPS and CASA are produced and managed by The National Radio Astronomy Observatory. The National Radio Astronomy Observatory is a facility of the National Science Foundation operated under cooperative agreement by Associated Universities, Inc. This work makes use of the ALFALFA survey, based on observations made with the Arecibo Observatory. The Arecibo Observatory is operated by SRI International under a cooperative agreement with the National Science Foundation (AST-1100968), and in alliance with Ana G. M\'endez-Universidad Metropolitana, and the Universities Space Research Association. We wish to acknowledge all members of the ALFALFA team for their work in making ALFALFA possible. This research has made use of the NASA/IPAC Extragalactic Database (NED), which is operated by the Jet Propulsion Laboratory, California Institute of Technology, under contract with the National Aeronautics and Space Administration. This research has made use of SAOImage DS9, developed by Smithsonian Astrophysical Observatory (http://hea-www.harvard.edu/RD/ds9/). Funding for the SDSS and SDSS-II was provided by the Alfred P. Sloan Foundation, the Participating Institutions, the National Science Foundation, the U.S. Department of Energy, the National Aeronautics and Space Administration, the Japanese Monbukagakusho, the Max Planck Society, and the Higher Education Funding Council for England. The SDSS Web Site is http://www.sdss.org/. GALEX is a NASA mission managed by the Jet Propulsion Laboratory. LN acknowledges STFC for the studentship grant and the LTA grant. LN acknowledges Haverford College for hosting her as a visiting scholar.

\bibliography{references}{}
\bibliographystyle{apj}

\end{document}